\documentstyle[psfig,rotating]{elsart}
\newcommand {\be}{\begin{equation}}
\newcommand {\ee}{\end{equation}}

\begin{document}
\bibliographystyle{unsrt}

\bigskip 
\begin{frontmatter}
\title{Nonlinear denoising of transient signals with application to event related potentials}

\author[MEB,ISKP]{A. Effern},
\author[MEB]{K. Lehnertz},
\author[PHYW]{T. Schreiber},
\author[MEB]{T. Grunwald},
\author[ISKP]{P. David}, 
\author[MEB]{C.E. Elger}

\address[MEB]{Department of Epileptology, University of Bonn,
  Sigmund-Freud Str. 25, \\53105 Bonn, Germany}
\address[ISKP]{Institute of Radiation and Nuclear Physics, University of Bonn,
  Nussallee 11-13, \\53115 Bonn, Germany}
\address[PHYW]{Department of Physics, University of Wuppertal, 
Gauss-Strasse 20, \\42097 Wuppertal}

\vspace{1cm}

\begin{abstract}
\baselineskip2.0em
We present a new wavelet based method for the denoising of {\it event related potentials}
(ERPs), employing techniques recently developed for the paradigm of deterministic chaotic systems.
The denoising scheme has been constructed to be appropriate for short and transient 
time sequences using circular state space embedding. Its effectiveness was successfully tested on simulated signals as well as on 
ERPs recorded from within a human brain. The method enables the study of individual ERPs 
against strong ongoing brain electrical activity.


\vspace{0.5cm}
\hspace{0.3cm}   87.22.-q
\hspace{0.3cm}   87.22.Jb

\vspace{0.5cm}

\end{abstract}
\end{frontmatter}

\baselineskip2.0em

\section{Introduction}

The {\it electroencephalogram} (EEG) reflects brain electrical activity
owing to both intrinsic dynamics and responses to external stimuli. To
examine pathways and time courses of information processing under specific
conditions, several experiments have been developed controlling sensory
inputs. Usually, well defined stimuli are repeatedly presented during
experimental sessions (e.g., simple tones, flashes, smells, or touches).
Each stimulus is assumed to induce synchronized neural activity in specific
regions of the brain, occurring as potential changes in the EEG. These {\it %
evoked potentials} (EPs) often exhibit multiphasic peak amplitudes within
the first hundred milliseconds after stimulus onset. They are specific for
different stages of information processing, thus giving access to both
temporal and spatial aspects of neural processes. Other classes of
experimental setups are used to investigate higher cognitive functions. For
example, subjects are requested to remember words, or perhaps they are asked
to respond to specific target stimuli, e.g. by pressing a button upon their
occurrence. The neural activity induced by this kind of stimulation also
leads to potential changes in the EEG. These {\it event related potentials}
(ERPs) can extend over a few seconds, exhibiting peak amplitudes mostly
later than EPs. Deviation of amplitudes and/or moment of occurrence
(latency) from those of normal EPs/ERPs are often associated with
dysfunction of the central nervous system and thus, are of high relevance
for diagnostic purposes.

As compared to the ongoing EEG, EPs and ERPs possess very low peak
amplitudes which, in most cases, are not recognizable by visual inspection.
Thus, to improve their low signal-to-noise ratio, EPs/ERPs are commonly
averaged (Figure 1), assuming synchronous, time-locked responses not
correlated with the ongoing EEG. In practice, however, these assumptions may
be inaccurate and, as a result of averaging, variations of EP/ERP latencies
and amplitudes are not accessed. In particular, short lasting alterations
which may provide relevant information about cognitive functions are
probably smoothed or even masked by the averaging process. Therefore,
investigators are interested in {\it single trial analysis}, that allows
extraction of reliable signal characteristics out of single EP/ERP sequences 
\cite{Lopes86}. In ref. \cite{Cerutti87} {\bf a}uto{\bf r}egressive models (%
{\it AR}) are adopted to EEG sequences recorded prior to stimulation in
order to subtract uncorrelated neural activity from ERPs. However, it is an
empirical fact, that external stimuli lead to
event-related-desynchronizaition of the ongoing EEG. Thus, the estimated
AR-model might be incorrect. The authors of \cite{Heinze84} applied {\bf a}%
uto{\bf r}egressive {\bf m}oving {\bf a}verage ({\it ARMA}) models to time
sequences which were a concatenation of several EP/ERP sequences. In the
case of short signal sequences, this led to better spectral estimations than
commonly achieved by periodograms. The main restriction is, however, that
investigated signals must be linear and stationary, which cannot be strictly
presumed for the EEG. In particular the high model order in comparison to
the signal length shows that AR- and ARMA-models are often inadequate for
EP/ERP analysis. Other methods have been developed to deal with the
nonstationary and transient character of EPs/ERPs. Woody \cite{Woody67}
introduced an iterative method for EP/ERP latency estimation based on common
averages. He determined the time instant of the best correlation between a
template (EP/ERP average) and single trials by shifting the latter in time.
This method corrects a possible latency variability of EPs/ERPs, but its
performance highly depends on the initial choice of templates. The {\it %
Wiener filter} \cite{Walter69,Doyle75}, on the other hand, uses spectral
estimation to reduce uncorrelated noise. This technique, however, is less
accurate for EPs/ERPs, because the time course of transient signals is lost
in the Fourier domain. Thus, DeWeerd \cite{DeWeerd81a,DeWeerd81b} introduced
a time adaptive Wiener filter, allowing better adjustment to signal
components of short duration. The paradigm of orthogonal wave packets ({\it %
wavelet transform}\footnote{%
Continuous wavelet transform: $w_{a,b}(\Psi ,x(t))=\frac{1}{\sqrt{\left|
a\right| }}\int_{-\infty }^{+\infty }x(t)\Psi (\frac{t-b}{a})dt$%
\par
$w$: wavelet coefficient, $a$: scaling parameter, $b$: translation
parameter, $x(t)$: time series, $\Psi $: {\it mother wavelet} function})
also follows this concept of adopted time-frequency decomposition. In
addition, the wavelet transform provide several useful properties which make
it preferable even for the analysis of transient signals \cite
{Burrus98,Chui92,Daubechies92}:

\begin{itemize}
\item  Wavelets can represent smooth functions as well as singularities.

\item  The basis functions are local which makes most coefficient based
algorithms to be naturally adapted to inhomogeneities in the function.

\item  They have the unconditional basis property to represent a variety of
functions implying that the wavelet basis is usually a reasonable choice
even if very little is known about the signal.

\item  Fast wavelet transform is computationally inexpensive of order $O(N)$%
, where $N$ denotes the number of sample points. In contrast, fast Fourier
transform (FFT) requires $O(Nlog(N))$.

\item  Nonlinear thresholding is nearly optimal for signal recovery.
\end{itemize}

For that reasons, wavelets became a popular tool for the analysis of brain
electrical activity \cite{Coifman96,Samar95,Schiff94a,Schiff94b}, especially
for denoising and classification of single trial EPs/ERPs. Donoho et al.\cite
{Donoho95} introduced a simple thresholding algorithm to reduce noise in the
wavelet domain requiring no assumptions about the time course of signals.
Nevertheless, high signal amplitudes are in need to distinguish between
noise and signal related wavelet coefficients in single trials. Bertrand et
al. \cite{Bertrand94} modified the original {\it a posteriori Wiener filter}
to find accurate filter settings. The authors emphasized better adoption to
transient signal components than can be achieved by corresponding techniques
in the frequency domain. However, due to the averaging process, this
technique runs the risk of choosing inadequate filter settings in the case
of a high latency variability. The same restriction is valid for
discriminant techniques applied e.g. by Bartink et al. \cite
{Bartink92a,Bartink92b}. Nevertheless, wavelet based methods enable a more
adequate treatment of transient signals than techniques applied in the
frequency domain. The question of accurate filter settings, however, is
still an unresolved problem.

To circumvent this problem, we introduce a new method for single trial
analysis of ERPs that neither assumes fully synchronized nor stationary ERP
sequences. The method is related to techniques already developed for the
paradigm of deterministic chaotic systems, using time delay embeddings of
signals for state space reconstruction and denoising \cite{Kantz97}.
Schreiber and Kaplan \cite{Schreiber95} demonstrated the accuracy of these
methods to reduce measurement noise in the human electrocardiogram (ECG).
Heart beats are also of transient character and exhibit relevant signal
components in a frequency range that compares to ERPs. Unfortunately, ERPs
are of shorter duration as compared to the ECG. Thus, in the case of high
dimensional time delay embedding (in the order of the signal length), we
cannot create a sufficient number of delay vectors for ERP sequences. To
circumvent this problem we reconstruct ERPs in state-space using circular
embeddings, that have turned out to be appropriate even for signal sequences
of short duration. In contrast to the nonlinear projection scheme described
in \cite{Kantz97}, we do not use {\it singular value decomposition} (SVD) to
determine clean signals in state space. The reason for this is threefold.
First, estimating relevant signal components using the inflexion of ordered
eigen-values is not always applicable to EEG because eigen-values may decay
almost linearly. In this case, an a priori restriction to a fixed embedding
dimension is in need, running the risk either to discard important signal
components or to remain noise of considerable amplitude if only little is
known about the signal. Second, SVD stresses the direction of highest
variances, so that transient signal components may be smoothed by
projection. Third, the number of signal related directions in state space
may alter locally, which is also not concerned by SVD. Instead we calculate
wavelet transforms of delay vectors and determine signal related components
by estimating variances separately for each state-space direction. Scaling
properties of wavelet bases allow very fast calculation as well as focusing
on specific frequency bands. To confirm the accuracy of our method, we apply
it to ERP-like test signals contaminated with different types of noise.
Afterwards, we give an example of reconstructed mesial temporal lobe P300
potentials, that were recorded from within the hippocampal formation of a
patient with focal epilepsy.

\section{Outline of the Method}

A time series may be contaminated by random noise allowing the measurement $%
y_{n}=x_{n}+\epsilon _{n}$. If the measured time series is purely
deterministic, it is restricted to a low-dimensional hyper-surface in state
space. For the transient signals we are concerned with here, we assume this
still to be valid. We hope to identify this direction and to correct $y_{n}$
by simply projecting it onto the subspace spanned by the clean data \cite
{Grassberger93,Schreiber95}.

Technically we realize projections onto noise free subspaces as follows. Let 
$Y=(y_{1},y_{2},\ldots ,y_{N})$ denote an observed time sequence. Time-delay
embedding of this sequence in a $m$-dimensional state space leads to state
space vectors ${\bf y}_{n}=(y_{n},\ldots ,y_{n-(m-1)\tau })$, where $\tau $
is an appropriate time delay. In an embedding space of dimension $m$ we
compute the discrete wavelet transform \cite{Daubechies92,Chui92,Burrus98}
of all delay vectors in a small neighborhood of a vector ${\bf y}_{n}$ we
want to correct. Let $r_{n,j}$ with $j=0,\ldots ,k$ denote the indices of
the k nearest neighbors of ${\bf y}_{n}$, and for ${\bf y}_{n}$ itself, i.e. 
$j=0$, and $r_{n,0}=n$. Thus, the first neighbor distances from ${\bf y}_{n}$
in increasing order are $d(Y)_{n}^{(1)}\equiv ||{\bf y}_{n}-{\bf y}%
_{r_{n,1}}||=\min_{r^{\prime }}||{\bf y}_{n}-{\bf y}_{r^{\prime }}||$, $%
d(Y)_{n}^{(2)}\equiv ||{\bf y}_{n}-{\bf y}_{r_{n,2}}||=\min_{{\ r^{\prime }}%
\neq r_{n,1}}||{\bf y}_{n}-{\bf y}_{r^{\prime }}||$, etc., where $||{\bf y}-%
{\bf y}^{\prime }||$ is the Euclidean distance in state space. Now the
important assumption is that the clean signal lies within a subspace of
dimension $d\ll m$, and that this subspace is spanned by only a few basis
functions in the wavelet domain. Let ${\bf w}_{r_{n,j}}$ denote the fast wavelet
transform \cite{Mallat89a,Mallat89b} of ${\bf y}_{r_{n,j}}$. Futhermore, let $%
C_{i}^{(k)}({\bf w}_{r_{n}})=\langle {\bf w}_{r_{n,j}}\rangle_i $ denote the $i^{th}$
component of the centre of mass of ${\bf w}_{r_{n}}$, and $\sigma _{n,i}^{2}$ the
corresponding variance. In the case of neighbors owing to the signal ({\it %
true neighbors}), we can expect the ratio $C_{i}^{(k)}({\bf w}_{r_{n}})/\sigma
_{n,i}^{2}$ to be higher in signal than in noise related directions. Thus, a
discrimination of noise and noise free components in state space is
possible. Let 
\begin{equation}
{\tilde{w}}_{n,i}=\left\{ 
\begin{array}{r@{\quad : \quad}l}
w_{n,i} & |C_{i}^{(k)}({\bf w}_{r_{n}})|\geq 2\lambda \frac{\sigma _{n,i}}{\sqrt{%
k+1}} \\ 
0 & $else$
\end{array}
\right. 
\end{equation}
define a shrinking condition to carry out projection onto a noise free
manifold \cite{Donoho95}. The parameter $\lambda $ denotes a thresholding
coefficient that depends on specific qualities of signal and noise. Inverse
fast wavelet transform of ${\tilde{{\bf w}}}_{n}$ provides a corrected vector in
state space, so that application of our projection scheme to all remaining
delay vectors ends up with a set of corrected vectors, out of which the
clean signal can be reconstructed.

\subsection{Extension to multiple signals of short length}

Let $Y_{l}=(y_{l,1},y_{l,2},\ldots ,y_{l,N})$ denote a short signal sequence
that is repeatedly recorded during an experiment, where $l=1,\ldots ,L$
orders the number of repetitions. A typical example may be ERP recordings,
where each $Y_{l}$ represents an EEG sequence following well defined
stimuli. Time-delay embeddings of these sequences can be written as ${\bf y}%
_{l,n}=(y_{l,n}\ldots ,y_{l,n-(m-1)\tau })$. To achieve a sufficient number
of delay vectors even for high embedding dimensions, we define circular
embeddings by 
\begin{equation}
{\bf y}_{l,n}=(y_{l,n},\ldots ,y_{l,1},y_{l,N},\ldots ,y_{l,N-(m-q)})\quad
\forall \quad n<m,
\end{equation}
so that all delay vectors with indices $1\leq n\leq N$ can be formed.
Circular embeddings are introduced as the most attractive choice to handle
the ends of sequences. Alternatives are (i) losing neighbors, (ii)
zeropadding, and (iii) shrinking the embedding dimension towards the ends.
However, discontinuities may occur at the edges, requiring some smoothing.
For each $Y_{l}$ we define the smoothed sequence as 
\begin{equation}
{\bf y}_{l,n,i}^{s}=\left\{ 
\begin{array}{l@{\quad : \quad}l}
y_{l,n,i}e^{-(\frac{q-i}{p})^{2}} & i<q \\ 
y_{l,n,i} & q\leq i\leq N-q \\ 
y_{l,n,i}e^{-(\frac{i-(N-q)}{p})^{2}} & i>N-q
\end{array}
\right.
\end{equation}
where $q$ defines the window width in sample points, $p$ the steepness of
exponential damping, and $i$ the time index. Time-delay embedding of several
short sequences leads to a filling of the state space, so that a sufficient
number of nearest neighbors can be found for each point.

\subsection{Parameter Selection}

Appropriate choice of parameters, in particular embedding dimension $m$,
time delay $\tau $, thresholding coefficient $\lambda $, as well as the
number of neighbors $k$ is important for accurate signal reconstruction in
state space. Several methods have been developed to estimate ``optimal''
parameters, depending on specific aspects of the given data (e.g., noise
level, type of noise, stationarity, etc.). These assume that the clean
signal is indeed low dimensional, an assumption we are not ready to make in
the case of ERPs. Thus, we approached the problem of ``optimal'' parameters
empirically.

Parameters $\tau $ and $m$ are not independent from each other. In
particular, high embedding dimensions allow small time-delays and vice
versa. We estimated ''optimal'' embedding dimensions and thresholding
coefficients on simulated data by varying $m$ and $\lambda $ for a fixed $%
\tau =1$. To allow fast wavelet transform, we chose $m$ to be a power of 2.

Repeated measurements, like in the case of EPs/ERPs, have a maximum number
of true neighbors which is given by $k_{max}=L$. In the case of identical
signals this is the best choice imaginable. However, real EPs/ERPs may alter
during experiments, and it seems more appropriate to use a maximum distance
true neighbors are assumed to be restricted to. We define this distance by 
\begin{equation}
d({\bf y})_{max}=\frac{\sqrt{2}}{LN}\sum_{l=1,n=1}^{L,N}{\ d({\bf y}%
)_{n,l}^{(L)}}
\end{equation}

\section{Model Data}

\subsection{Generating test signals and noise}

To demonstrate the effectiveness of our denoising technique and to estimate
accurate values for $m$, $\lambda $, and $L$, we applied it to EP/ERP-like
test signals contaminated with white noise and in-band noise. The latter was
generated using phase randomized surrogates of the original signal \cite
{Theiler92}. Test signals consisted of 256 sample points and were a
concatenation of several Gaussian functions with different standard
deviations and amplitudes. To simulate EPs/ERPs not fully synchronized with
stimulus onset, test signals were shifted randomly in time (normally
deviated, std. dev.: $20$ sample points, max. shift: $40$ sample points).
Since even fast components of the test signal extended over several sample
points, a minimum embedding dimensions $m=16$ was required to cover any
significant fraction of the signal. The highest embedding dimension was
bounded by the length of signal sequences and the number of embedded trials,
thus allowing a maximum of $m=256$. However, if the embedding dimension is $%
m=N$, neighborhood is not longer defined by local characteristics, and we
can expect denoised signals to be smoothed in the case of multiple time
varying components.

\subsection{Denoising of test signals}

Let $X_{l}=(x_{l,1},x_{l,2},\ldots ,x_{l,N})$ denote the $l^{th}$ signal
sequence of a repeated measurement, $Y_{l}=(y_{l,1},y_{l,2},\ldots ,y_{l,N})$
the noise contaminated sequence, and $\tilde{Y}_{l}=(\tilde{y}_{l,1},\tilde{y%
}_{l,2},\ldots ,\tilde{y}_{l,N})$ the corresponding result of denoising.
Then 
\begin{equation}
{\bf r}=\frac{1}{L}\sum_{l=1}^{L}\sqrt{\frac{{(Y_{l}-X_{l})^{2}}}{{(\tilde{Y}_{l}-X_{l})^{2}}}}
\end{equation}
defines the {noise reduction factor} which quantifies signal improvement
owing to the filter process.

We determined {\bf r} for test signals contaminated with white noise, using
noise amplitudes ranging from 25\% - 150\%, and embedding dimensions ranging
from 16 - 128 (Figure 2a, Figure 3). Five repetitions for each parameter
configuration were calculated using 5 embedded trials each. In the case of $%
\lambda \leq 2$, the {\it noise reduction factor} was quite stable against
changes of noise levels but depended on embedding dimension $m$ and
thresholding coefficient $\lambda $. Best performance was achieved for $%
1.0\leq \lambda \leq 2.0$ (${\bf r}_{max}^{m=128,\lambda =2.0}=4.7$). In the
case of $\lambda >4.0$, most signal components were rejected, and as a
result, the {\it noise reduction factor} ${\bf r}$ increased linearly with
noise levels, as expected.
Figure 2b and Figure 4 depict effects of denoising of 5 test signals
contaminated with in-band noise. In comparison to white noise the
performance decreased, but nevertheless, enabled satisfactory denoising for $%
0.5\leq \lambda \leq 1.0$ (${\bf r}_{max}^{m=128,\lambda =1.0}=1.6$). Within
this range, the {\it noise reduction factor} ${\bf r}$ depended weakly on
noise levels. Note that the embedding dimension must be sufficiently high ($%
m=128$) to find true neighbors.

In order to simulate EPs/ERPs with several time-varying components, we used
5 test signals which were again a concatenation of different Gaussian
functions, each, however, randomly shifted in time (Figure 2c and Figure 5).
In contrast to test signals with time fixed components, ''optimal''
embedding dimension depended on the thresholding coefficient $\lambda $.
Higher values of $\lambda $ required lower embedding dimensions and vice
versa. Best results were achieved for $0.5\leq \lambda \leq 2.0$ (${\bf r}%
_{max}^{m=128,\lambda =1.0}=3.2$).

Even for high noise levels, the proposed denoising scheme preserved finer
structures of original test signals in all simulations. Moreover, the
reconstructed sequences were closer to the test signals than the
corresponding averages, especially for time varying signals. Power spectra
showed that denoising took part in all frequency bands and was quite
different from common low-, or band-pass filtering. Simulation indicated
that ''optimal'' values of the thresholding coefficient were in the range $%
0.5\leq \lambda \leq 2.0$. Best embedding dimension was found to be $m=128$,
since the ongoing background EEG can be assumed to be in-band with ERPs. The
filter performance was quite stable against the number of embedded
sequences, at least for $L = 5, 10, 20$.

\section{Real data}

\subsection{Data Acquisition}

We analyzed event related potentials recorded intracerebrally in patients
with pharmacoresistent focal epilepsy \cite{Grunwald99}.
Electroencephalographic signals were recorded from bilateral electrodes
implanted along the longitudinal axis of the hippocampus. Each electrode
carried 10 cylindrical contacts of nickel-chromium alloy with a length of
2.5 mm and an intercontact distance of 4 mm. Signals were referenced to
linked mastoids, amplified with a bandpass filter setting of 0.05 - 85.00 Hz
(12dB/oct.) and, after 12 bit A/D conversion, continuously written to a hard
disk using a sampling interval of $5760\mu $s. Stimulus related epochs
spanning 1480 ms (256 sample points) including a 200 ms pre-stimulus
baseline were extracted from recorded data. The mean of the pre-stimulus
baseline was used to correct possible amplitude shifts of the following ERP
epoch.

In a {\it visual odd-ball paradigm }60 rare (letter $<x>$, targets) and 240
frequent stimuli (letter $<o>$, distractors) were randomly presented on a
computer monitor once every $1200\pm 200ms$ (duration: 100 ms, probability
of occurrence: 1 ($<x>$) : 5 ($<o>$)). Patients were asked to press a button
upon each rare target stimulus. This pseudo-random presentation of rare
stimuli in combination with the required response is known to elicit the
mesial temporal lobe (MTL) P300 potential in recordings from within the
hippocampal formation \cite{Puce89} (cf. Figure 1).

\subsection{Results}

By simulation, we estimated a range in which "optimal" parameters of the
filter can be expected. However, the quality of denoising ERP sequences
could not be estimated, because the clean signal was not known a priori. A
rough estimation of filter performance was only possible by a comparison to
ERP averages. Taking into account results of simulation as well as ERP
averages, we estimated $\lambda=0.6$ and $m=128$ to be the best
configuration.

Based on the empirical fact that specific ERP components exhibit peak
amplitudes within a narrow time range related to stimulus onset, we defined
a maximum allowed time jitter of $\pm 20$ sample points ($\approx 116ms$)
true neighbors are assumed to be restricted to. This accelerated the
calculation time and avoided false nearest neighbors. Figure 6 depicts
several ERPs recorded from different electrode contacts within the
hippocampal formation. The number of embedded sequences was chosen as $L=8$.
Comparing averages, we can expect that the filter extracted the most
relevant MTL-P300 components. Even for low amplitude signals reconstruction
was possible, exhibiting higher amplitudes in single trial data than in
averages. As corresponding power spectra show, the 50 Hz power line was
reduced but not eliminated after filtering. Especially low amplitude signals
showed artifacts based on the 50 Hz power line.

\section{Conclusion}

In this study, we introduced a new wavelet based method for nonlinear noise
reduction of single trial EPs/ERPs. We employed advantages of methods
developed for the paradigm of deterministic chaotic systems, that allowed
denoising of short and time variant EP/ERP sequences without assuming fully
synchronized or stationary EEG.

Denoising via wavelet shrinkage does not require a priori assumptions about
constrained dimensions, as is usually required for other techniques (e.g.,
singular value decomposition). Besides, it is more straight forward using
thresholds depending on means and variances rather than initial assumptions
about constrained embedding dimensions. Moreover, the local calculation of
thresholds in state space enables focusing on specific frequency scales,
which may be advantageous in order to extract signal components located
within both narrow frequency bands and narrow time windows.

Extension of our denoising scheme to other types of signals seems to be
possible, however, demands further investigations, since ''optimal'' filter
parameters highly depend on signal characteristics. In addition, the {\it %
noise reduction factor} ${\bf r}$ does not consider all imaginable features
of signals investigators are possibly interested in, so that other measures
may be more advantageous in specific cases.

So far, we have not considered effects of smoothing the edges of signal
sequences. But since delay vectors as well as corresponding wavelet
coefficients hold information locally, we can assume artifacts to be also
constrained to the edges which we were not interested in.

In conclusion, the proposed denoising scheme represents a powerful noise
reduction technique for transient signals of short duration, like ERPs.

\vspace{0.3cm} {\bf Acknowledgements}

This work is supported by the Deutsche Forschungsgemeinschaft (grant. no. EL
122 / 4-2. ).

We thank G. Widman, W. Burr, K. Sternickel, and C. Rieke for
fruitful discussions.

\newpage {\bf Figure captions:} \vspace{1cm}

Fig. 1: Examples of averaged ERPs recorded along the longitudinal axis of the
hippocampal formation in a patient with epilepsy. Randomized presentation of 
{\it target} and {\it standard} stimuli is known to elicit the mesial
temporal lobe P300, a negative deflection peaking at about 500 ms after
stimulus onset (cf. Sect. 4.1 for more details). Letters (a),
(b), and (c) indicate recordings used for single trial analysis (cf. Figure
6).

Fig. 2: Results of denoising test signals. Parts a) and b): contamination
with white noise and in-band noise. Part c): time varying signal components
and white noise contamination (see text for more details). Five calculations
for each parameter configuration have been executed to determine standard
deviations.

Fig. 3: Nonlinear denoising applied to white noise contaminated test signals
(5 sequences embedded, each 256 sample points, randomly shifted in time
(std. dev.: 20 sample points, max. shift: 40 sample points), noise amplitude
75\%, $m=128$, $\tau =1$, $\lambda =1.5$). Power spectra in arbitrary units.
For state space plots we used a time delay of 25 sample points.

\bigskip

Fig. 4: Same as Figure 3 but for in-band noise and $\lambda =0.75$.

\bigskip

Fig. 5: Same as Figure 3 but for Gaussian functions each randomly shifted in
time and $\lambda =0.75$. 

Fig. 6: Examples of denoised MTL-P300 potentials (cf. Figure 1). 
Power spectra in arbitrary units. For state
space plots we used a time delay of 25 sample points. 

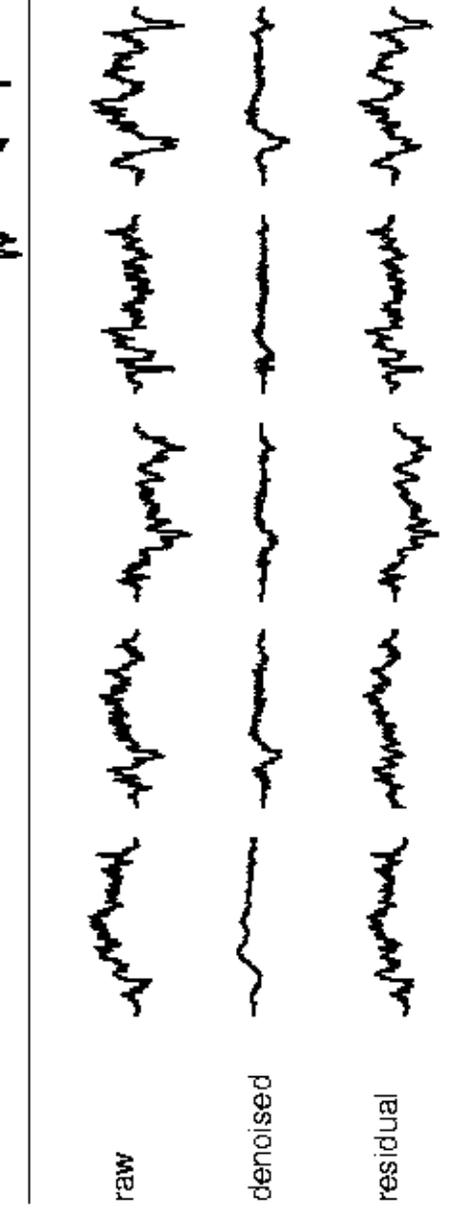
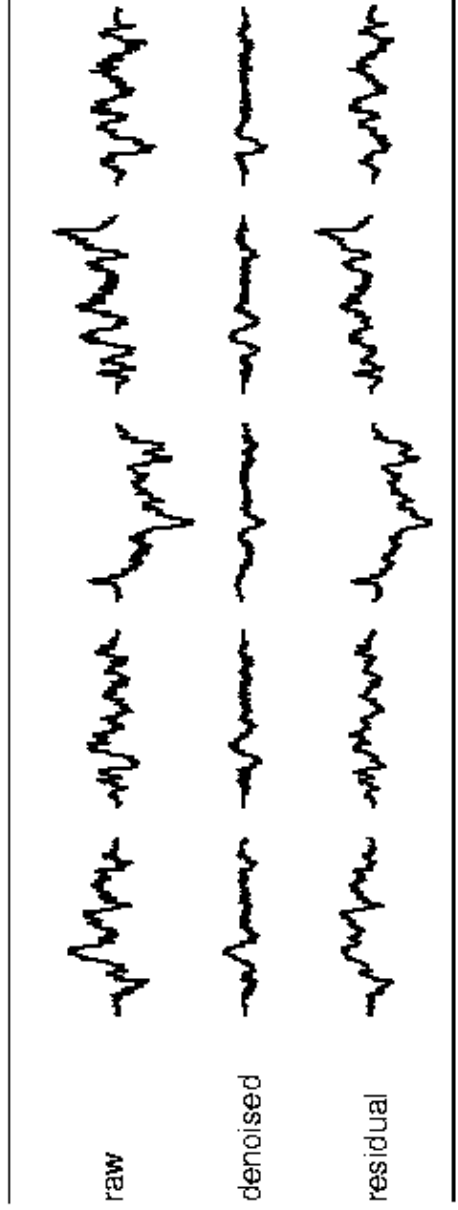
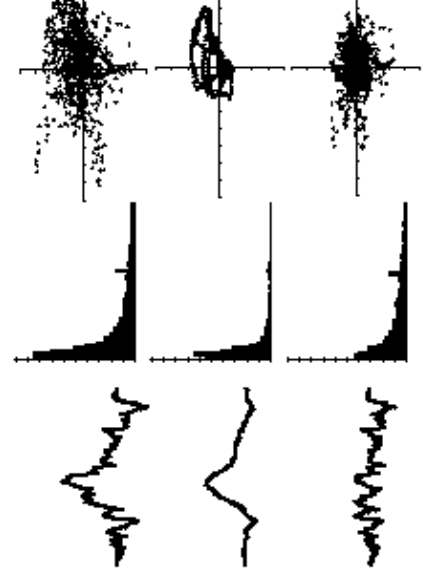
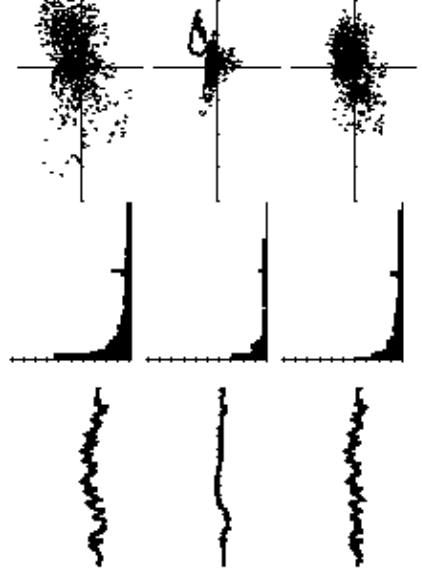
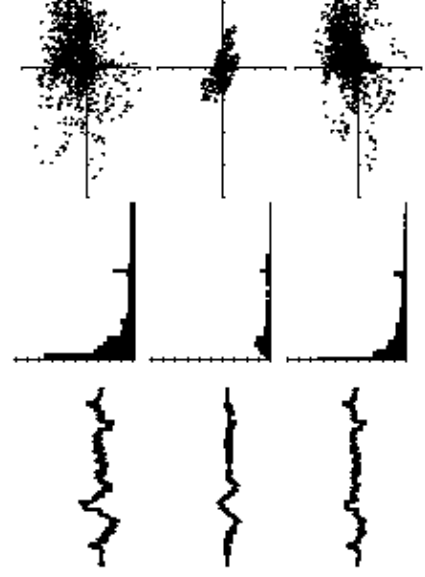
\end{document}